\documentclass[letterpaper,twocolumn,english,superscriptaddress,letterpaper, nofootinbib]{revtex4}

\usepackage[]{fontenc}
\usepackage[latin9]{inputenc}
\usepackage{textcomp}
\usepackage{amsmath}
\usepackage{graphicx}
\usepackage{amssymb}
\usepackage{bbm}
\usepackage{dsfont}
\usepackage{newtxmath}
\usepackage{dcolumn}
\usepackage{babel}
\usepackage{color}
\usepackage[normalem]{ulem}

\makeatletter

\pdfpageheight\paperheight
\pdfpagewidth\paperwidth

\newcommand{\lyxmathsym}[1]{\ifmmode\begingroup\def\b@ld{bold}
  \text{\ifx\math@version\b@ld\bfseries\fi#1}\endgroup\else#1\fi}

\@ifundefined{textcolor}{}
{%
 \definecolor{BLACK}{gray}{0}
 \definecolor{WHITE}{gray}{1}
 \definecolor{RED}{rgb}{1,0,0}
 \definecolor{GREEN}{rgb}{0,1,0}
 \definecolor{BLUE}{rgb}{0,0,1}
 \definecolor{CYAN}{cmyk}{1,0,0,0}
 \definecolor{MAGENTA}{cmyk}{0,1,0,0}
 \definecolor{YELLOW}{cmyk}{0,0,1,0}
 }

\newcommand{\ms}{Mn$_{3}$Sn}

\makeatother

\begin{document}

\title{Imprinting electrically switchable scalar spin chirality by anisotropic strain in a Kagome antiferromagnet} 

\author{Debjoty Paul}
\altaffiliation{These authors contributed equally to this work.}
\affiliation{Department of Condensed Matter Physics and Materials Science, Tata Institute of Fundamental Research, Homi Bhabha Road, Mumbai 400005, India}

\author{Shivesh Yadav}
\altaffiliation{These authors contributed equally to this work.}
\affiliation{Department of Condensed Matter Physics and Materials Science, Tata Institute of Fundamental Research, Homi Bhabha Road, Mumbai 400005, India}

\author{Shikhar Gupta}
\affiliation{Department of Condensed Matter Physics and Materials Science, Tata Institute of Fundamental Research, Homi Bhabha Road, Mumbai 400005, India}

\author{Bikash Patra}
\affiliation{Department of Condensed Matter Physics and Materials Science, Tata Institute of Fundamental Research, Homi Bhabha Road, Mumbai 400005, India}

\author{Nilesh Kulkarni}
\affiliation{Department of Condensed Matter Physics and Materials Science, Tata Institute of Fundamental Research, Homi Bhabha Road, Mumbai 400005, India}

\author{Debashis Mondal}
\affiliation{Department of Condensed Matter Physics and Materials Science, Tata Institute of Fundamental Research, Homi Bhabha Road, Mumbai 400005, India}

\author{Kaushal Gavankar}
\affiliation{Department of Condensed Matter Physics and Materials Science, Tata Institute of Fundamental Research, Homi Bhabha Road, Mumbai 400005, India}

\author{Sourav K. Sahu}
\affiliation{Department of Condensed Matter Physics and Materials Science, Tata Institute of Fundamental Research, Homi Bhabha Road, Mumbai 400005, India}
\affiliation{School of Physical Sciences, National Institute of Science Education and Research, An OCC of Homi Bhabha National Institute, Jatni 752050, India}

\author{Biswarup Satpati}
\affiliation{Surface Physics \& Material Science Division, Saha Institute of Nuclear Physics, A CI of Homi Bhabha National Institute, 1/AF Bidhannagar, Kolkata 700064, India}

\author{Bahadur Singh}
\affiliation{Department of Condensed Matter Physics and Materials Science, Tata Institute of Fundamental Research, Homi Bhabha Road, Mumbai 400005, India}

\author{Owen Benton}
\affiliation{School of Physical and Chemical Sciences, Queen Mary University of London, London E1 4NS, United Kingdom}

\author{Shouvik Chatterjee}
\email[Authors to whom correspondence should be addressed: ]{shouvik.chatterjee@tifr.res.in}
\affiliation{Department of Condensed Matter Physics and Materials Science, Tata Institute of Fundamental Research, Homi Bhabha Road, Mumbai 400005, India}

\begin{abstract}

Topological chiral antiferromagnets, such as Mn$_{3}$Sn, are emerging as promising materials for next-generation spintronic devices due to their intrinsic transport properties linked to exotic magnetic configurations. Here, we demonstrate that anisotropic strain in Mn$_{3}$Sn thin films offers a novel approach to manipulate the magnetic ground state, unlocking new functionalities in this material. Anisotropic strain reduces the point group symmetry of the manganese (Mn) Kagome triangles from $C_{3v}$ to $C_{1}$, significantly altering the energy landscape of the magnetic states in Mn$_{3}$Sn. This symmetry reduction enables even a tiny in-plane Dzyaloshinskii-Moriya (DM) interaction to induce canting of the Mn spins out of the Kagome plane. The modified magnetic ground state introduces a finite scalar spin chirality and results in a significant Berry phase in momentum space. Consequently, a large anomalous Hall effect emerges in the Kagome plane at room temperature - an effect that is absent in the bulk material. Moreover, this two-fold degenerate magnetic state enables the creation of multiple-stable, non-volatile anomalous Hall resistance (AHR) memory states. These states are field-stable and can be controlled by thermal assisted current-induced magnetization switching requiring modest current densities and small bias fields, thereby offering a compelling new functionality in Mn$_{3}$Sn for spintronic applications.

\end{abstract}

\maketitle

%%%%%%%%%%%%%%%%%%%%%%%%%%%%%% Introduction%%%%%%%%%%%%%%%%%%%%%%%%%%%%%% 
\section{Introduction}

A Kagome lattice consists of corner sharing triangles, where a delicate interplay between geometrical frustration, magneto-crystalline anisotropies, and competing interactions can give rise to unconventional magnetic ground states including ordered states that are chiral \cite{wang2023quantum,wang2024topological}. It can also host a non-trivial band structure in the momentum space, which is often intertwined with the magnetic order \cite{ye2018massive,kang2020dirac,kang2020topological}. Such chiral Kagome antiferromagnets exhibits unique intrinsic transport properties, which make them attractive for potential applications in next-generation logic, memory, and computing technologies. One such compound is \ms\/ that has been reported to have time-reversal symmetry breaking Weyl points \cite{yang2017topological,kubler2018weyl,kuroda2017evidence,chen2021anomalous} and exhibits a large anomalous Hall effect with negligible magnetization at room temperature \cite{nakatsuji2015large}. \ms\/ has a hexagonal Ni$_{3}$Sn-type crystal structure (space group P6$_{3}$/mmc) that can be viewed as a bi-layer stacking of the Kagome planes along the $c$-axis/[0001] direction, as shown in Fig.~1(a). The manganese (Mn) atoms form a two-dimensional Kagome lattice in the (0001)/$a$-$b$ plane with the tin (Sn) atoms at the centre of the hexagon. In the bulk, \ms\/ has a non-collinear inverse triangular spin structure over a wide temperature range between 50 K and the N\'eel temperature of $T_{N}$ = 420 K \cite{ohmori1987spin}. The Mn spins lie within the Kagome plane and rotate 120\textdegree\/ anti-clockwise along a clockwise circulation of the Kagome triangle formed by the Mn atoms (see Fig.~1(a)) \cite{tomiyoshi1982magnetic,brown1990determination,kren1975study}. Magnetocrystalline anisotropy in \ms\/ cants the Mn spins slightly towards the Mn-Sn easy axis within the Kagome plane resulting in a small uncompensated moment \cite{tomiyoshi1982magnetic,brown1990determination,nagamiya1982triangular,sandratskii1996role}. Although magnetization is tiny, it shows a large anomalous Hall effect (AHE) at room temperature when the magnetic field is applied within the Kagome plane, which stems from intrinsic Berry curvature in the momentum space \cite{nakatsuji2015large,li2023field}. However, such a coplanar spin configuration preserves the combined symmetry $TM_{z}$, where $T$ is the time reversal and $M_{z}$ is the mirror symmetry with the (0001)/$a$-$b$ plane being the mirror plane, and the vertical glide symmetry $\tilde{M_{x}}$ = $\{M_{x} \vert \frac{c}{2}\}$, where $M_{x}$ is the mirror symmetry with the (11$\bar{2}$0)/\textit{a-c} plane being the mirror plane.  This constrains Berry curvature ($\Omega_{xy}$) to be $\Omega_{xy}(k_{x},k_{y},k_{z}) = -\Omega_{xy}(-k_{x},-k_{y},k_{z})$. Accordingly, intrinsic Berry curvature contribution to the anomalous Hall conductivity (AHC) ($\sigma^{z}_{xy}$) in the Kagome ($x$-$y$) plane is identically zero, in accordance with experimental observation, both in single crystals and thin films \cite{nakatsuji2015large,li2023field,ikeda2020fabrication,taylor2020anomalous}.

In this study, we demonstrate that the application of anisotropic strain, combined with an in-plane Dzyaloshinskii-Moriya (DM) interaction induces a canting of the Mn spins out of the Kagome plane in \ms\/. This spin canting leads to the emergence of a non-zero scalar spin chirality ($\chi = S_{i} \cdot (S_{j} \times S_{k}) \neq 0$) at room temperature. Importantly, the out-of-plane spin canting breaks both the $TM_{z}$ and $\tilde{M_{x}}$ symmetries, resulting in the appearance of a finite Berry curvature in the Kagome plane ($\sigma^{z}_{xy}$). Consequently, we observe the emergence of a large anomalous Hall conductivity (AHC, $\sigma^{z}_{xy}$) in the Kagome plane. This AHC, previously absent in both bulk single crystals and thin films of \ms\/, is a novel finding not observed in earlier studies \cite{nakatsuji2015large, li2023field, ikeda2020fabrication, taylor2020anomalous}. Furthermore, we show that this anomalous Hall response can be fully switched at room temperature by applying an electrical current pulse, which induces thermal-assisted switching. The two-fold degenerate nature of the newly induced magnetic ground state allows for the stabilization of different domain configurations during the switching process. This feature enables the realization of electrically controlled, multiple-stable, non-volatile anomalous Hall resistance (AHR) memory states, unlocking a new functionality in \ms\/ that can be useful for potential applications in spintronics and neuromorphic computing.

%%%%%%%%%%%%%%%%%%Results & Discussion%%%%%%%%%%%%%%%%%%%%%%%%%%%%%% 

\section{Results and Discussion}

%%%%%%%%%%%%%%%%%%Growth & Transport%%%%%%%%%%%%%%%%%%%%%%%%%%%%%% 

\subsection{Anisotropic strain in epitaxial Mn$_{3}$Sn (0001) thin films}

Epitaxial, (0001) out-of-plane oriented thin films of \ms\/ were fabricated by magnetron sputtering on $c$-plane sapphire substrates using tantalum (Ta), having the alpha ($\alpha$) phase, as a buffer layer. Ta grows epitaxially on $c$-plane sapphire with Ta(110) planes oriented along the out-of-plane direction and having three domains in the in-plane direction that are rotated at 120 degrees from each other (see Fig.~S1 in the Supplementary Information)\cite{jones2023grain,alegria2023two}. This provides a quasi-six-fold symmetric template for epitaxial integration of \ms\/(0001) thin films\cite{suppl}. Epitaxial relationship in our thin films is \ms\/ ($10\bar{1}0$)[$0001$]$||$Ta($100$)[$110$]$||$Al$_{2}$O$_{3}$($11\bar{2}0$)[$0001$], determined from x-ray diffraction measurements. However, symmetry and lattice mismatch between Ta(110) and \ms\/(0001) surfaces imparts an anisotropic strain on \ms\/ atomic layers, which is borne out from our ab-initio calculations, shown in Fig.~1(b). We estimate the average lattice parameters of the Mn Kagome triangle in \ms\/ by measuring the reciprocal lattice vectors ($q_{\parallel}, q_{z}$) of \ms\/\{$20\bar{2}1$\} family of diffraction spots, shown in Fig.~1(d) and Fig.~S3 in the Supplementary Information \cite{suppl}. The anisotropic strain results in a reduction of the $C_{3v}$ point group symmetry of the Kagome triangles to $C_{1}$ (see Fig.~1(b)). The inter-atomic distances between the nearest neighboring Mn atoms in the Kagome triangle are now different, with the experimentally obtained average values of 2.845, 2.856, and 2.860\AA\/, respectively, as shown in Fig.~1(a), in contrast to the uniform value of 2.838\AA\/ reported in the bulk \cite{sung2018magnetic}. The out-of-plane lattice parameter, $c$ = 4.538\AA\/ in our thin films is similar to the reported value of 4.536\AA\/ in the bulk \cite{sung2018magnetic}. High-resolution transmission electron microscopy (HR-TEM) on these thin films also show similar lattice parameters \cite{suppl}. For further details on the estimation of the lattice parameters, please refer to the Supplementary Information \cite{suppl}. The \ms\//Ta epitaxial heterostructures were capped with either AlO$_{x}$ or Ta/AlO$_{x}$ protective layers before exposing them to the ambient atmosphere.

 Atomic layers of Ta has a strong spin-orbit coupling and breaks the inversion symmetry along the $c$-axis at the \ms\//Ta interface. It, therefore, can stabilize a strong in-plane interfacial DM interaction \cite{dzyaloshinsky1958thermodynamic,moriya1960anisotropic,yang2015anatomy} in the \ms\//Ta thin film heterostructures. Furthermore, the large lattice and symmetry mismatch at the heteroepitaxial interface leads to an appreciable dislocation density in \ms\/. These dislocations give rise to local strain and strain gradients, which breaks the local inversion symmetry in the bulk of \ms\/ atomic layers. This is directly observed in the strain analysis of our thin films, shown in the Supplementary Information (see section S3 and Fig.~S5)\cite{suppl}. Moreover, our ab-initio calculations also reveal inversion symmetry breaking in the relaxed \ms\/ structure in \ms\//Ta heterostructure, shown in Fig.~1(b). Inversion symmetry breaking in the bulk gives rise to a finite bulk DM interaction in \ms\/ atomic layers when grown on Ta, which is otherwise absent in pristine \ms\/. The reduction of point group symmetry in \ms\/ due to anisotropic strain (see Fig.~1(a)) and in-plane DM interaction, both bulk and interfacial, plays an important role in stabilizing non-trivial magnetic ground state in these thin films, which will be discussed in the following sections. 
 
 Ta also serves as an ideal buffer layer because it is stable against both Mn and Sn. This is established directly from the energy-dispersive x-ray spectroscopy (EdS) data of our thin films and from ab-initio calculations (please see section 4 and Figs. S6, S7 in the Supplementary Information)\cite{suppl}. This stability enables high-temperature annealing ($\geq$ 450\textdegree\/ C) of \ms\/, which is crucial for achieving epitaxial thin films with superior structural and electrical properties. This is in contrast to platinum (Pt), where interfacial reactions occur between \ms\/ and Pt at temperatures above 350\textdegree\/C \cite{tsai2021large}. Additional characterization of the thin film heterostructures can be found in the Supplementary Information \cite{suppl}.

\subsection{Anomalous Hall effect in Mn$_{3}$Sn thin films with anisotropic strain}

Having established the structural properties of the thin film heterostructures, we present their electrical and magnetic properties in Fig.~2. We observe a large AHC at room temperature when the measurements are done in the Kagome plane with the magnetic field applied perpendicular to it.  This is in sharp contrast to the earlier reports on both bulk single crystals \cite{nakatsuji2015large,li2023field} as well as epitaxial \ms\/(0001) thin films 
\cite{ikeda2020fabrication,taylor2020anomalous}, where no AHC ($\sigma^{z}_{xy}$) is observed in the Kagome plane at room temperature. In those studies, no evidence for spontaneous magnetic moment was found when the magnetic field was applied perpendicular to the Kagome plane. This confirms that the co-planar spin configuration of \ms\/, which preserves $TM_{z}$ and $\tilde{M_{x}}$  symmetries enforced $\sigma^{z}_{xy}$ to be zero in those experiments.  In our case, we observe a tiny spontaneous magnetic moment indicative of a finite canting of the Mn spins out of the Kagome plane. This induces finite scalar spin chirality and explicitly breaks the $TM_{z}$ and $\tilde{M_{x}}$ symmetries allowing the emergence of AHC ($\sigma^{z}_{xy}$), as observed in our measurements. The magnetic moment was found to be $\approx$ $9.3$ and $16.5$ $m\mu_{B}$ per Mn for 90 nm thick \ms\/ films when capped with AlO$_{x}$ (Ta(11 nm)/ \ms\/(90 nm)/ AlO$_{x}$(8 nm)) and Ta/AlO$_{x}$ (Ta(11 nm)/ \ms\/(90 nm)/ Ta(11 nm)/ AlO$_{x}$(8 nm)), respectively, as shown in Fig.~2(c),(d). The corresponding AHCs are $\sigma^{z}_{xy}$(H=0) $\approx$ $27.7$ and $33.7$ $\Omega^{-1}$cm$^{-1}$, respectively (see Fig.~2(a),(b)). The carrier concentration, which is estimated from the ordinary Hall effect, and longitudinal resistivities in \ms\/ at room temperature are found to be similar in both kinds of heterostructures. We estimated a carrier concentration of $5.9 \times 10^{21}$/cc and $6.1 \times 10^{21}$/cc and longitudinal resistivities of 268.64 and 267.35 $\mu$Ohm-cm for the thin films capped with AlO$_{x}$ and Ta/AlO$_{x}$, respectively. This establishes that \ms\/ atomic layers are not affected by the presence of adjacent Ta layers providing additional evidence of a stable hetero-interface between \ms\/ and Ta with minimal chemical intermixing. 
We note that the magnitudes of the AHC($\sigma^{z}_{xy}$) in the Kagome plane of our thin films heterostructures are large, similar to what has been observed in bulk single crystals and thin films. However, in those cases AHC has only been observed in planes perpendicular to the Kagome plane. This, therefore, indicates a different magnetic ground state in our thin films compared to what has been reported before. Such a large AHC cannot be explained by the tiny magnetization ($M_{z}$), shown in Figs.~2(c,d),\cite{nakatsuji2015large} but points to a prominent role of scalar spin chirality and induced Berry curvature in the momentum space.

\subsection{Anisotropic strain and in-plane Dzyaloshinskii-Moriya interaction induces scalar spin chirality in Mn$_{3}$Sn}

To understand how an out-of-plane spin canting is stabilized in \ms\//Ta heterostructures, we evaluate the ordered magnetic ground state of the Kagome lattice, which respects all the relevant symmetries. The full crystal structure of Mn$_3$Sn is a 3D layered structure, with substantial coupling between the Kagome planes. However, making use of the observed ${\bf q}=0$ order, one can use the fact that spins of the same sub-lattice in neighboring planes takes on the same value, and reduce the problem to that of a single 
layer with renormalized coefficients\cite{liu2017anomalous}.

The  Hamiltonian can then be written as a sum over all Kagome triangles

\begin{equation}
    H=\sum_{\triangle} H_{\triangle}
\end{equation}
where
\begin{multline}
    H_{\triangle} = (J\mathbf{S_{0}}.\mathbf{S_{1}} + \mathbf{D_{01}}.(\mathbf{S_{0}}\times\mathbf{S_{1}})) + 
  (J\mathbf{S_{1}}.\mathbf{S_{2}} + \mathbf{D_{12}}.(\mathbf{S_{1}}\times\mathbf{S_{2}})) +\\
    (J\mathbf{S_{2}}.\mathbf{S_{0}} + \mathbf{D_{20}}.(\mathbf{S_{2}}\times\mathbf{S_{0}})) +
\frac{1}{2}\sum_{i}K(\hat{\mathbf{n_{i}}}.\mathbf{S_{i}})^{2}
\label{eq:H_unstrained}
\end{multline}
and $J$, $\mathbf{D_{ij}}$, and $K$ are the Heisenberg exchange, DM interaction, and easy-axis single ion anisotropy, respectively, where the local easy axis $\mathbf{\hat{n_{i}}}$ is oriented from site $i$ to the center of the Kagome hexagons. The factor $\frac{1}{2}$ on the single-ion term is included to account for double counting when the total energy of the lattice is calculated by summing over all Kagome triangles. There is a hierarchy in the energy scale: $J$ $>$ $D$ $>$ $K$, which is typical for $3d$ transition metal ions such as Mn \cite{liu2017anomalous}. $S_{i}$, $i\in 0,1,2$ denotes the Mn spins on three Mn sublattices in \ms\/, as shown in Fig.~3(b). In the absence of strain all Mn-Mn bonds in the Kagome triangles are of same length and related to each other by $C_{3}$ rotation symmetry (Fig.~1(a)). Therefore, the Heisenberg exchange interaction is expected to be of identical strength for each of the three bonds, $J_{ij} = J \forall \langle i,j\rangle$, as used in eqn.~2. The symmetry allowed DM interaction, $\mathbf{D_{ij}}$, can be written as $\mathbf{D_{ij}}$ = D$_{z}$$\mathbf{\hat{z}}$ + D$_{\parallel}$($\mathbf{\hat{z}\times\hat{e_{ij}}}$), where $\mathbf{\hat{e_{ij}}}$ is the unit vector oriented from site $i$ to site $j$, shown schematically in Fig.~S8 in the Supplementary Information\cite{suppl}. The reflection symmetry of the Kagome planes in the ideal crystal structure of \ms\/ forces $D_{\parallel}$ to be identically zero. However, in  \ms\//Ta heterostructures, the hetero-epitaxial interface breaks the inversion symmetry along (0001), as does strain relaxation in the bulk of \ms\/ atomic layers, as described in the previous section. Hence, DM interaction in \ms\//Ta heterostructures can have both interfacial and bulk $D_{\parallel}$ components. Following ref. \cite{liu2017anomalous} we set $\frac{D_{z}}{J}$ and $\frac{K}{J}$ as 0.1 and -0.03, respectively (note the different sign convention for $\frac{K}{J}$). For experimentally feasible values of $\frac{D_{\parallel}}{J}$ ($\frac{D_{\parallel}}{J}$ $\leq$ $\pm$0.2), the most stable spin structure is found to be the coplanar inverse triangular spin structure, which is six-fold degenerate (labelled E-coplanar$_{6}$, following the notation in ref.~\cite{benton2021ordered}) and does not possess scalar spin chirality ($\chi=0$, see Fig.~3(a)). Our theoretical results, therefore, are in accordance with prior experimental reports on bulk single crystals and epitaxial (0001) thin films \cite{nakatsuji2015large, li2023field, ikeda2020fabrication, taylor2020anomalous, sung2018magnetic, park2018magnetic}. Hence, the presence of in-plane DM interaction alone is not sufficient to induce scalar spin chirality in \ms\//Ta heterostructures. 

Next, we incorporate anisotropic strain, as observed in \ms\//Ta heterostructures, where all three Mn-Mn bonds alter their lengths and become inequivalent, reducing the point group symmetry of the Kagome triangles to $C_{1}$ (see Fig.~1(a)). To account for this, we introduce factors $\alpha_{ij}$, which renormalize the strength of interactions on each of the three bonds. The simplification of the Hamiltonian down to a single Kagome layer described above eqn.~(\ref{eq:H_unstrained}) remains valid in the presence of strain, provided that the strain is not so strong as to break the assumption of ${\bf q}=0$ order. This is even true in the presence of inhomogenous strain, observed in \ms\//Ta heterostructures. We also note that the strength of inversion symmetry breaking in these heterostructures, and hence that of $D_{\parallel}$, is expected to be stronger closer to the interface. In such cases, the parameters in the Hamiltonian, such as $D_{\parallel}$, should then be considered as averaged effective parameters. 

Therefore, under anisotropic strain the Hamiltonian can be written as:
\begin{multline}
    H_{\triangle} = \alpha_{01}(J\mathbf{S_{0}}.\mathbf{S_{1}} + \mathbf{D_{01}}.(\mathbf{S_{0}}\times\mathbf{S_{1}})) + 
    \alpha_{12}(J\mathbf{S_{1}}.\mathbf{S_{2}} + \mathbf{D_{12}}.(\mathbf{S_{1}}\times\mathbf{S_{2}})) +\\
    \alpha_{20}(J\mathbf{S_{2}}.\mathbf{S_{0}} + \mathbf{D_{20}}.(\mathbf{S_{2}}\times\mathbf{S_{0}})) +
\frac{1}{2}\sum_{i}K(\hat{\mathbf{n_{i}}}.\mathbf{S_{i}})^{2}
\label{eq:Htri}
\end{multline}
The factors $\alpha_{ij}$ are \textit{a priori} unknown, which we will relate to the change in average bond lengths, which are known. We assume here for simplicity that the Heisenberg and DM interactions are modified by the same factor. We do this by assuming that the exchange interactions are purely a function of the bond distance, i.e. for Heisenberg exchange $J = J(l)$ and
\begin{equation}
    \alpha_{ij} = \frac{J(l_{ij})}{J(l_{0})}
\end{equation}
We then Taylor expand the function $J(l)$ around the unstrained bond length $l_{0}$:
\begin{multline}
    \alpha_{ij} \approx \frac{J(l_{0}) + (l_{ij} - l_{0})(\frac{dJ}{dl})_{l = l_{0}}}{J(l_{0})} = 1+\frac{l_{0}}{J(l_{0})}\left(\frac{dJ}{dl}\right)_{l=l_{0}}\left(\frac{l_{ij} - l_{0}}{l_{0}}\right)\\
    = 1 - \lambda\left(\frac{l_{ij} - l_{0}}{l_{0}}\right)
\end{multline}
where the dimensionless parameter $\lambda \equiv  - \frac{l_{0}}{J(l_{0})}\left(\frac{dJ}{dl}\right)_{l=l_{0}}$ expresses the sensitivity of the exchange interactions to small changes in bond length. A minus sign is included in the definition of $\lambda$ because of the intuitive expectation that exchange interactions decrease with increasing $l$. Our calculations predict that anisotropic strain, parametrized by $\lambda$ ($\lambda$ = 0 corresponds to the unstrained case), stabilizes a non-coplanar ground state ($\chi$ $\neq$ 0) under any finite in-plane DM interaction, as shown in the phase diagram in Fig.~3(a). In such cases the ground state is two-fold degenerate and is denoted as E-noncoplanar$_{2}$. Following a recent neutron scattering experiment \cite{park2018magnetic}, which experimentally parametrized the exchange interactions in \ms\/ with small differences in bond length in the Kagome plane, we estimate $\lambda \approx -39$. This suggests that exchange interactions in \ms\/ can not only become stronger with increasing bond length ($\lambda$ \textless\/ 0) but the change can be quite large in magnitude. Representative non-coplanar magnetic ground states corresponding to both $\lambda = +39$ and $\lambda = -39$ with a net out-of-plane magnetic moment ($M_{z}$ $\approx$ 9.3 $m\mu_{B}$/Mn) similar to what is measured in our thin films, are shown in Fig.~3(b). For both the cases, the value of $D_{\parallel}/J$ was tuned to reproduce the same out-of-plane magnetization, giving  $D_{\parallel}/J=0.14$ for $\lambda=39$ and $D_{\parallel}/J=0.19$ for $\lambda=-39$. We wish to emphasize that any non-zero $\lambda$, i.e. anisotropic strain in the Kagome lattice in presence of a finite in-plane DM interaction will stabilize a non-coplanar ground state, as is evident from the phase diagram presented in Fig.~3(a). We do not attempt a precise fit of the Hamiltonian parameters here, as it would be under-constrained, but the parameter sets used for calculations are illustrative of the behavior across parameter space. Further details can be found in the Methods and in the Supplementary Information\cite{suppl}.

We incorporated the E-noncoplanar$_{2}$ magnetic structure shown in Fig.~3(b) in our first-principles calculations and calculated the resulting band structure. In Fig.~\ref{fig:Calculation}(c), we present the band structure for the coplanar spin structure showing a nearly gapless nodal line along the $K-H$ line without any Berry curvature. Considering the non-coplanar spin configurations (E-noncoplanar$_{2}$), corresponding to $\lambda = +39$ and $\lambda = -39$, the nodal line becomes fully gapped (Fig.~\ref{fig:Calculation}(d)). Importantly, E-noncoplanar$_{2}$ ground state breaks both the combined $TM_{z}$ and the vertical glide mirror $\tilde{M_{x}}$ = $\{M_{x} \vert \frac{c}{2}\}$, which opens a large gap at nodal-band crossings. This results in a large Berry curvature $\Omega_{xy}$, which is smeared out in energy and has a finite contribution at the Fermi energy, in agreement with our experimental observation of large AHC. The calculated value of $\sigma^{z}_{xy}$ is 1.37 ($\lambda = +39$) and -0.71 ($\lambda = -39$) S/cm at the Fermi level, which goes up to 30 ($\lambda = +39$) and 31 ($\lambda = -39$) S/cm at 40 and 38 meV below the Fermi level, respectively (see Fig.~3(e)). Notably, relaxing the magnetic structure in our DFT calculations, where substrate is not included, always chooses a coplanar magnetic configuration ($\chi$ = 0) as the ground state, with a large difference in energy with any non-coplanar spin configuration ($\chi$ $\neq$ 0). This is in agreement with our theory calculations, which predicts a coplanar inverse triangular spin structure (E-coplanar$_{6}$) for all $\lambda$ values when in-plane DM interaction ($D_{\parallel})$, which is not present in the ideal crystal structure of \ms\/, is zero. However, when \ms\//Ta heterostructure is considered explicitly in our DFT calculations, we observe inversion symmetry breaking in the relaxed \ms\/ structure and the stabilization of non-coplanar magnetic states in \ms\/ (See Fig.~1(b) and Fig.S6 in the Supplementary Information\cite{suppl}) Our work, therefore, underscores the importance of both in-plane DM interaction and anisotropic strain in stabilizing a non-coplanar ground state in the Kagome lattice. 

We emphasize that the non-coplanar ground state is stabilized for any finite anisotropic strain and in-plane DM interaction, and is independent of the actual values of these parameters. However, for any finite anisotropic strain ($\lambda \neq 0$), a stronger DM interaction leads to an enhanced out-of-plane spin canting (see Fig.~3(a)), which is expected to result in a larger Berry curvature contribution. This is also borne out in our experiments. Samples which are capped with Ta shows a larger out-of-plane spin canting and AHC (see Fig.~2(b),(d)), where DM interaction strength is expected to be stronger due to an additional \ms\//Ta hetero-interface. Additionally, we note that thin films of \ms\/ synthesized on a 4$d$ metal with a weak spin-orbit coupling (ruthenium), does not exhibit anomalous Hall conductivity in the Kagome plane\cite{taylor2020anomalous}. This highlights the importance of a strong spin-orbit coupled 5$d$ metal, Ta, and its contribution towards generating a large in-plane DM interaction in \ms\//Ta heterostructures.

\subsection{Thermal assisted electrical switching of anomalous Hall effect and realization of multiple-stable memory states}

Having understood the origin of anomalous Hall effect in \ms\/(0001)/Ta heterostructures, we now investigate the possibility of switching the non-coplanar magnetic order with electric current. Electrical switching of the coplanar inverse triangular spin structure leading to a reversal of sign of the AHE in \ms\/ has already been reported, where the Mn spins rotates within the Kagome plane \cite{tsai2020electrical,higo2022perpendicular,takeuchi2021chiral}. The non-collinear inverse triangular spin configuration of \ms\/ has been shown to have cluster magnetic octupole as the order parameter \cite{suzuki2017cluster}. It has been claimed that spin-orbit torque (SOT) from an adjacent heavy-metal layer under the application of an in-plane magnetic field can deterministically switch the direction of the octupole moment, similar to spin-orbit torque switching of magnetization in a ferromagnet \cite{manchon2019current,chernyshov2009evidence,miron2010current}, which in turn results in the switching of the AHE \cite{tsai2020electrical,higo2022perpendicular}. We note that although the reported spin diffusion length in \ms\/ is very short ($\approx$ 1 nm)\cite{muduli2019evaluation}, spin-orbit torque driven switching has been observed in thick \ms\/ films ($>$ 90 nm), where recent reports suggest a possible role of thermally driven demagnetization on application of charge current in seeding the antiferromagnetic order under a spin-orbit field.\cite{pal2022setting,krishnaswamy2022time}

In our thin films not only the magnetic ground state is different, as already discussed, so is its switching behavior under the application of electric pulse. Although Ta layer in \ms\//Ta can generate appreciable spin current, we did not observe any evidence of spin-orbit torque driven switching behavior in our thin films. In contrast to expectation from switching by a spin-orbit field, the electrically induced switching in our case depends only on the magnitude, but not on the sign of the applied current. Furthermore, in our case, deterministic switching of anomalous Hall effect was observed under a tiny bias magnetic field applied perpendicular to the Kagome plane. Dependence of the observed switching behavior on both current pulse amplitude and pulse width is shown in Fig.~4. In both the cases, switching was observed only when the transient temperature of the device went above the N\'eel temperature of \ms\/ ($T_{N}$ $\approx$ 420K), which clearly establishes that it is driven by thermal effects. Under an applied electric pulse, the sample undergoes demagnetization as the temperature rises above the N\'eel temperature. A small but finite cooling field lifts the two-fold degeneracy of E-noncoplanar$_{2}$ magnetic ground state and sets the net magnetization of the nucleated domains along the direction of the field. This, in turn, results in a switching of the sign of AHE on reversing the direction of the cooling field. 
Due to the two-fold degeneracy of the E-noncoplanar$_{2}$ ground state, applying an electric pulse in zero magnetic field leads to the formation of magnetic domains with equal net magnetization along the +z and -z directions after demagnetization.
In such cases, the net anomalous Hall effect in the sample cancels out to zero, as observed in our experiment (see Fig.~4(a)). Further details are provided in the Methods and in the Supplementary Information \cite{suppl}. The unique magnetic ground state and thermally driven switching behavior in anisotropic strained \ms\//Ta thin film heterostructures allow access to a number of different non-volatile anomalous Hall resistance(AHR) states, which correspond to different domain configurations. Such AHR states can be accessed by the application of electrical pulse under different bias fields, as shown in Fig.~5. A symmetry-breaking bias field of $\approx$ 0.13 Tesla is found to be sufficient for Ta/\ms\//Ta/AlO$_{x}$ heterostructures to completely saturate the domain structure. At saturation, the out-of-plane magnetization ($M_{z}$) of all the domains are aligned along the applied bias field that results in a full switching of the AHE on application of current. Similar behavior is also observed in Ta/\ms\//AlO$_{x}$ heterostructures \cite{suppl}. Application of different bias fields, which are less than the saturation field, stabilizes unique domain structure post electric pulse driven demagnetization in \ms\/. Under those conditions the proportion of domains aligned along $+z$ and $-z$ directions depends on the applied bias field during the switching process and results in the observation of different anomalous Hall resistances. This allows us to achieve multiple AHR memory states in \ms\/ that are tunable by electric current. Moreover, the change in AHR is linear over a wide magnetic bias field range of $\pm$200 Oe as shown in Fig.~5(b). The switching behavior is symmetric with respect to the polarity of the bias field, as evidenced from the resistance loops shown in Fig.~5(c). The resistance loops at different bias fields are obtained as follows: (i) At first, an electric pulse is applied at a positive bias field. (ii) Then, the bias field is reversed (iii) The same electric pulse is applied at a reversed bias field, which reverses the sign of AHR (R$_{xy}$). (iv) The applied field is reversed again, bringing it back to its original value. (v) Application of electric pulse brings the AHR back to its position at the end of step (i). These steps are shown in Fig. 5(c). The AHR retraces the same resistance loop on repetition of these five steps indicating the stability of the switching process. The AHR memory states are also fairly stable against magnetic field fluctuations. Once a particular memory state is achieved, we show the evolution of such a state under the application of magnetic field in Fig.~5(d). We choose the criterion for field stability as the field range over which the resistance stays within 2$\%$ of the full range ($\Delta$ R$_{xy}$ = $2/100\times0.21$ $\Omega$ = $0.0042$ $\Omega$), which allows us to define 50 unique AHR memory states. We find that all such states are stable within a field range of at least $\approx$ 1769 Oe, indicative of good resilience against field fluctuations. 

The estimated current density in \ms\/ atomic layers required for switching the AHE in \ms\/ in both Ta/\ms\//AlO$_{x}$ and Ta/\ms\//Ta/AlO$_{x}$ heterostructures is $\approx$ $1.5 \times 10^{10}$ A/m$^{2}$, which is  an order of magnitude lower than spin-orbit torque switching of magnetic octupole in \ms\/ thin films \cite{higo2022perpendicular} and spin-orbit torque driven switching in typical ferromagnets. Distinct switching behavior in \ms\//Ta heterostructures provides additional evidence for the stabilization of non-coplanar magnetic ground state in this system.

%%%%%%%%%%%%%%%%%%Conclusion%%%%%%%%%%%%%%%%%%%%%%%%%%%%%% 

\section{Conclusion}

In summary, we demonstrated that anisotropic strain and in-plane interfacial DM interaction enable the imprinting of scalar spin chirality in the magnetic ground state of a Kagome lattice. This leads to significant Berry curvature effects from gapped nodal lines in momentum space, resulting in a pronounced AHE in the Kagome plane at room temperature. Furthermore, we have showed that applying an electric pulse can induce thermally assisted switching of the AHE, allowing access to multiple non-volatile, field-stable AHR memory states - an uncommon feature in antiferromagnets, which has been missing in topological Kagome antiferromagnets such as \ms\/. We emphasize that AHR memory states are realized due to the unique two-fold generate E-noncoplanar$_{2}$ ground state in anisotropically strained \ms\//Ta heterostructures, which is absent in bulk \ms\/. Coupled with the large AHE at room temperature, this new functionality of electrically controlling these memory states positions Kagome antiferromagnets as promising candidates for spintronics and neuromorphic computing applications. Additionally, finite scalar spin chirality may stabilize non-trivial chiral ground states, including skyrmion phases, which will be explored in future research. While our results were demonstrated in the prototypical system of \ms\/, these concepts should be applicable to other Kagome lattice systems where thin film engineering techniques such as anisotropic strain and inversion symmetry breaking at hetero-epitaxial interfaces can be employed to stabilize non-trivial magnetic ground states and to realize novel functionalities in these compounds.
%%%%%%%%%%%%%%%%%%Acknowledgment%%%%%%%%%%%%%%%%%%%%%%%%%%%%%% 

\section*{Acknowledgments}

We thank Devendra Buddhikot, Ganesh Jangam, and Bhagyashree Chalke for technical assistance. We acknowledge the Department of Science and Technology (DST), SERB grant SRG/2021/000414 and Department of Atomic Energy (DAE) of the Government of India (12-R$\&$D-TFR-5.10-0100) for support. We thank Dr. Sunil Ojha and Dr. G. R. Umapathy for  Rutherford backscattering spectrometry (RBS) measurements on \ms\/ thin films at Inter-University Accelerator Centre (IUAC), Delhi. We also acknowledge the use of the laser writing facility at National Nano Fabrication Centre (NNFC), IISc.

%%%%%%%%%%%%%%%%%%Author Contributions%%%%%%%%%%%%%%%%%%%%%%%%%%%%%% 

\section*{Author Contributions}

D.P. and S.Y. contributed equally to this work. S.C. conceived the project and was responsible for its overall execution. S.C., D.P., and S.Y. planned the experiments. D.P. with assistance from N.K., D.M., and S.G. fabricated the thin films and performed structural characterization. B.Sat. prepared the thin film lamellae and performed TEM characterization. D.P. with assistance from S.G. performed the strain analysis using the TEM images. D.P. and S.Y. fabricated the Hall bar devices. S.Y. and D.P. with assistance from S.G. and S.K.S. performed the electrical and magnetometry measurements. D.P., with assistance from S.Y., N.K, K.G, S.K.S., and S.C., analyzed the data. O.B. performed the numerical calculations. B.P. and B.S. performed the ab-initio calculations. O.B., B.P., and B.S. provided theoretical explanation. D.P. and S.C. wrote the manuscript with inputs from O.B., B.P., and B.S. All authors discussed the results and commented on the manuscript.

\section*{Data Availability}

The data that supports the findings of this study are available from the corresponding author upon reasonable request.
%%%%%%%%%%%%%%%%%%Materials and Methods%%%%%%%%%%%%%%%%%%%%%%%%%%%%%% 

\section*{Competing Interests}
The authors declare no competing interests.

\section*{Materials and Methods}

\subsection{Thin film preparation and characterization}

High purity targets of Mn (99.99\%) \& Sn (99.99\%) were co-sputtered using RF magnetron sputtering to synthesize epitaxial thin film of \ms\/. Before deposition of \ms\/, Ta was sputtered at 600\textdegree\/C on epitaxial grade c-plane sapphire substrates using RF magnetron sputtering in the same chamber at a 10 mTorr pressure followed by an annealing period of 10 mins at the same temperature. The sample was subsequently cooled to 300\textdegree\/C temperature to synthesize Mn$_{3}$Sn films at a growth rate of $\approx$ 1.25 \AA/s. Following which either an AlO$_{x}$ protective capping layer or Ta(11 nm) was deposited on the sample at room temperature. An \textit{in situ} post-annealing treatment of one-hour duration was performed at 450\textdegree\/C followed by a slow cooling rate of $\approx$ 1.5\textdegree\/C per minute. The high-temperature post-annealing step was found to be essential for the realization \ms\/ thin films of good structural and phase quality. For Ta/Mn$_{3}$Sn/Ta samples, finally an AlO$_{x}$ protective layer was deposited at room temperature to prevent oxidation. Rutherford backscattering spectrometry (RBS) was used to examine the stoichiometry of \ms\/ in thin film heterostructures. The measurements were done using 2MeV alpha particles at the RBS beamline (Pelletron Accelerator RBS-AMS Systems (PARAS)) at Inter-University Accelerator Centre (IUAC), Delhi. The fits to the obtained data were performed using RUMP software. 

The crystal structure of thin film heterostructures was evaluated using x-ray diffraction (XRD) and high-resolution transmission electron microscopy (HR-TEM). XRD data was collected using a Rigaku SmartLab X-ray diffractometer with a 9 kW rotating anode with Cu K$\alpha$ radiation. Cross-sectional TEM samples were prepared using conventional mechanical thinning followed by Ar ion milling using precision ion polishing system (PIPS, GATAN Inc.) at an energy of 4.0 keV and cleaning at 1.5 keV. HR-TEM images were taken using a FEI Tecnai G2 F30-ST microscope operated at 300 keV. The HR-TEM images were analyzed using the open-source software package Strain++\cite{strain}, implementing geometric phase analysis (GPA)\cite{hytch1998quantitative} , to extract quantitative strain maps in \ms\/ atomic layers, shown in the Supplementary Information\cite{suppl}.\\

\subsection{Magnetotransport and electrical switching measurements}

For magnetotransport and electrical switching measurements, Hall bars of width 50$\mu$m were fabricated using standard optical photolithography techniques. The Hall resistivity and the magnetoresistance were measured in a commercial PPMS (Quantum Design) using low-frequency a.c. lock-in technique. For the switching measurements, the electrical pulses were applied using a Keithley 6221 current source, where the Hall voltage before and after the application of the pulse was measured using a Keithley 2182A nanovoltmeter. The transient longitudinal voltage was collected using a fast NI-DAQ-6210 with a sampling rate 250KS/s, from where transient temperature during the application of pulse was estimated \cite{suppl}. The magnetometry measurements were performed in a commercial SQUID-VSM magnetometer (Quantum Design).

\subsection{Theory calculations}

The effect of strain and DM interaction on the spin configuration of Mn$_3$Sn was calculated in a classical approximation, where the $S=3/2$ moments are treated as classical vectors of fixed length. The Hamiltonian is given as a sum over kagome triangles as described in eqns.~1 and 2. For such a Hamiltonian, the classical energy is known to be minimized by translationally invariant states \cite{benton2021ordered}, so it is sufficient to minimize the energy on a single triangle and then tile the resulting configuration over the whole system. Optimal configurations on a single triangle were found by numerically minimizing $H_{\triangle}$ with respect to the spin directions ${\bf S}_{0,1,2}$, shown in Fig.~3(b). 

First-principles calculations were performed within the framework of density functional theory using the projector augmented wave potential with the Vienna Ab initio Simulation Package (VASP)~\cite{kresse1996efficient, kresse1999from}. Generalized gradient approximation with the Perdew-Burke-Ernzerhof parametrization~\cite{perdew1996generalized} was used to include exchange-correlation effects. The self-consistent relativistic calculations were performed with a plane wave cutoff energy of 400 eV and a $\Gamma$ centered $10\times 10 \times 12$ k-mesh for Brillouin zone sampling. Material-specific tight-binding Hamiltonian was constructed by generating Wannier functions derived from Mn-{\it{s}}, {\it{d}}, and Sn-{\it{p}} states~\cite{mostofia2014an}. The anomalous Hall conductivities were calculated using a dense $350 \times 350 \times 350$ points grid with the WannierBerry code~\cite{tsirkin2021high}.

%\bibliography{main}

%\bibliographystyle{Science}

%%%%%%%%%%%%%%%%%%Bibliography%%%%%%%%%%%%%%%%%%%%%%%%%%%%%% 
%%%%%%%%%%%%%%%%%%Bibliography%%%%%%%%%%%%%%%%%%%%%%%%%%%%%% 

%%%%%%%%%%%%Figures%%%%%%%%%%%%%%%%%%%%%%%%%%%%%%%%%%%%%%%%%%%%%%%%%%%%%%%%%

%%%%%%%%%%%%Figure1%%%%%%%%%%%%%%%%%%%%%%%%%%%%%%%%%%%%%%%%%%%%%%%%%%%%%%%%%%%%%%%%%%

\begin{figure*}
\includegraphics[width=1\textwidth]{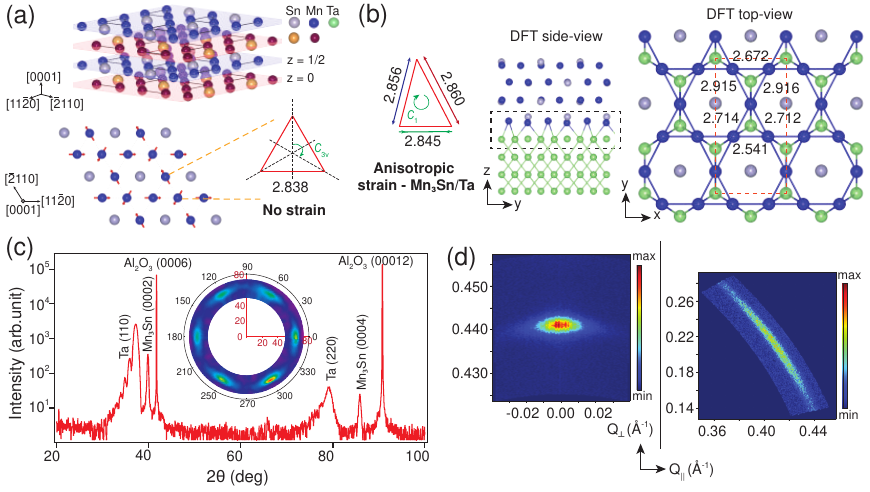}
\caption{\textbf{Anisotropic strain in \ms\//Ta heterostructures.} (a) Three-dimensional crystal structure of \ms\/, which consists of bilayer (AB) stacking of Kagome planes. Top view of the pristine Kagome plane and Mn Kagome triangles with $C_{3v}$ point group symmetry. (b) (left) Top view of the Mn Kagome triangle with anisotropic strain with a reduced symmetry ($C_{1}$) and experimentally measured lattice parameters. Side view of the relaxed structure of Ta(110)/\ms\/(0001) heterostructure considered in the density functional theory (DFT) calculations (middle) and a top view of the highlighted rectangular region (right), revealing a reduction of the $C_{3v}$ symmetry of the Kagome triangles to $C_{1}$ as well as lifting of the inversion symmetry of the pristine \ms\/ crystal structure. (c) Out-of-plane $\theta$-2$\theta$ XRD scan of \ms\//Ta heterostructures. Inset shows the pole figure plot of \ms\/ \{$20\bar{2}1$\} family of diffraction peaks showing a six-fold symmetry, establishing the epitaxial nature of \ms\/. (d) RSM plots of ($0002$) and ($20\bar{2}1$) Bragg peaks. The in-plane lattice parameters are estimated by measuring the $q$ vectors of the \{$20\bar{2}1$\} family of Bragg peaks\cite{suppl}. }
\label{fig:Structure}
\end{figure*}

%%%%%%%%%%%%Figure2%%%%%%%%%%%%%%%%%%%%%%%%%%%%%%%%%%%%%%%%%%%%%%%%%%%%%%%%%%%%%%%%%%
%%%%%%%%%%%%%%%Fig.2%%%%%%%%%%%%%%%%%%%%%%%%%%%%%%%%
\begin{figure}
\includegraphics[width=1\columnwidth]{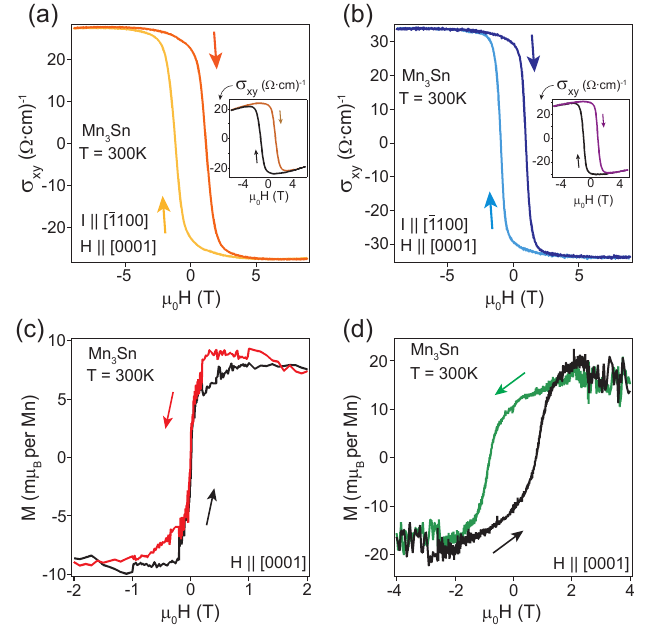}
\caption{\textbf{Anomalous Hall effect in \ms\//Ta heterostructures.} Anomalous Hall conductivity at 300K of (a) Ta(11nm)/ \ms\/(90nm)/ AlO$_{x}$(8nm) and (b) Ta(11nm)/ \ms\/(90nm)/ Ta(11nm)/ AlO$_{x}$(8nm) thin film heterostructures synthesized on c-plane sapphire substrates. The arrows indicate the direction of magnetic field sweeps, which is applied perpendicular to the Kagome plane (along [$0001$]) and the current is applied along [$\bar{1}100$]. Insets show corresponding Hall conductivity as a function of magnetic field. Magnetization as a function of magnetic field at 300K for (c) Ta(11nm)/ \ms\/(90nm)/ AlO$_{x}$(8nm) and (d) Ta(11nm)/ \ms\/(90nm)/ Ta(11nm)/ AlO$_{x}$(8nm) thin film heterostructures with the magnetic field applied perpendicular to the Kagome plane. A linear diamagnetic background has been subtracted in both the cases.}
\label{fig:Hall}
\end{figure}

%%%%%%%%%Fig.3%%%%%%%%%%%%%%%%%%%%%%%%%%%%%%%%%%%%%%
\begin{figure*}
\includegraphics[width=1\textwidth]{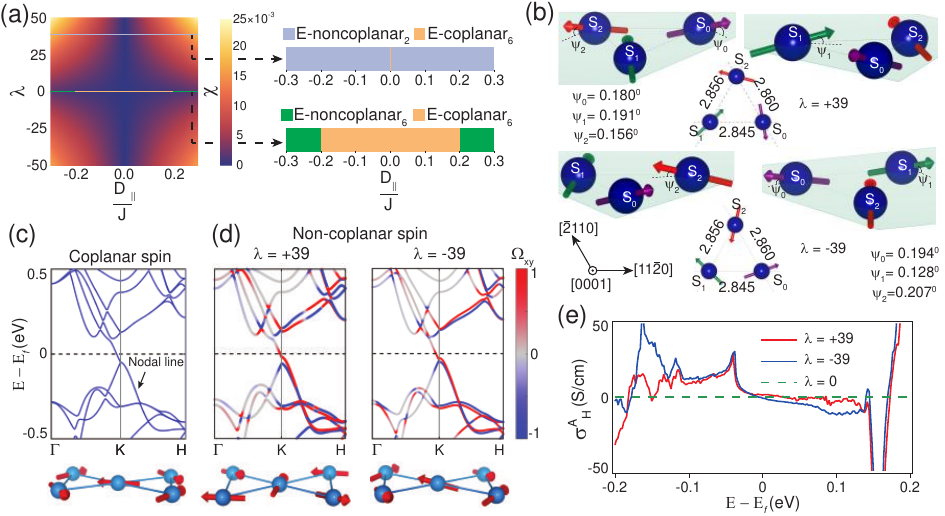}
\caption{\textbf{Non-coplanar spin structure in \ms\//Ta heterostructures.} (a) Phase diagram of the magnetic ground state as a function of anisotropic strain and in-plane DM interaction. Scalar spin chirality($\chi$) of the magnetic ground state as a function of $\lambda$ and $D_{\parallel}/J$ is shown on the left (all parameters defined in the text). Phase diagram as a function of $D_{\parallel}/J$ for two specific $\lambda$ values viz. $\lambda$ = 0, and $\lambda$ = +39 are shown on the right. (b) Illustration of non-coplanar magnetic ground states (E-noncolpanar$_{2}$) having an out-of-plane magnetic moment of $M_{z}$ $\approx$ 9.3 $m\mu B$/Mn corresponding to $\lambda$ = +39 and $\lambda$ = -39, as described in the text. Band structure obtained with (c) coplanar inverse triangular spin structure (E-coplanar$_{6}$, $\lambda$ = 0) and (d) non-coplanar (E-noncoplanar$_{2}$) spin configurations corresponding to $\lambda$ = +39 and $\lambda$ = -39, as shown in (b). The bottom panels show the spin-configurations and the color bar represents the Berry curvature $\Omega_{xy}$. (e) Calculated anomalous Hall conductivities (AHC)as a function of binding energy for $\lambda$ = 0, +39, and -39. Note that AHC is non-zero for $\lambda$ $\neq$ 0, and is identically zero otherwise. }
\label{fig:Calculation}
\end{figure*}
%%%%%%%%%%%%%%%%%%%%%%%%%%%%%%%%%%%%%%%%%%%%%%%%%%

%%%%%%%%%%%%%%%Fig.4%%%%%%%%%%%%%%%%%%%%%%%%%%%%%%%%
\begin{figure*}
\includegraphics[width=1\textwidth]{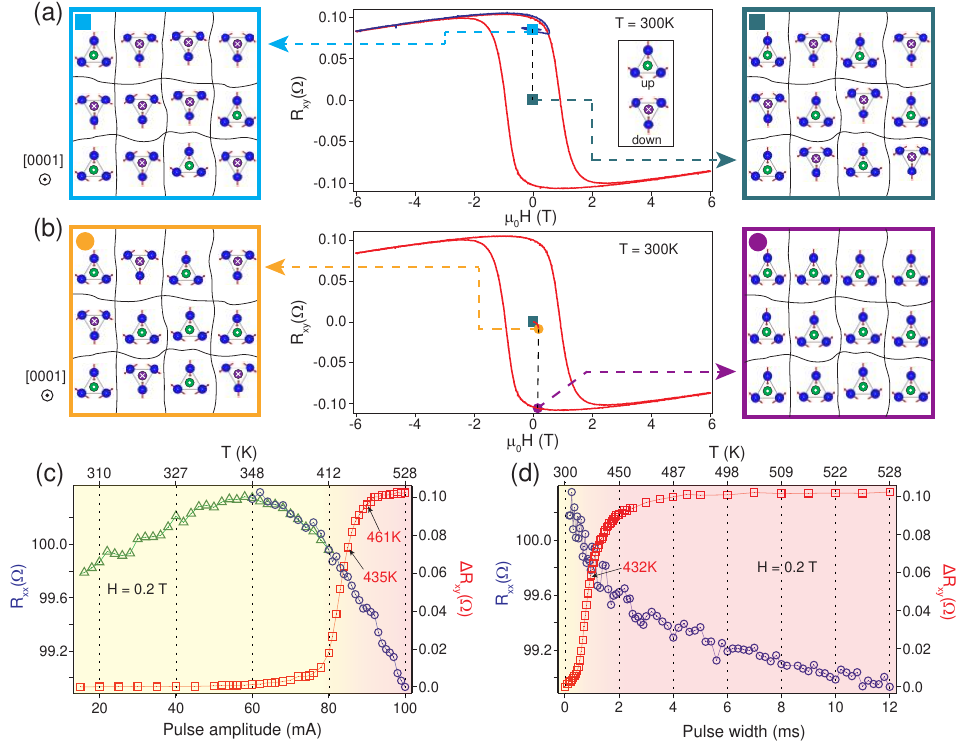}
\caption{ \textbf{Thermal assisted switching of anomalous Hall effect in \ms\/.} Switching of anomalous Hall resistance (a) under zero magnetic field and (b) under a finite magnetic field of $+0.2$ T in Ta(11nm)/\ms\/(90nm)/Ta(11nm)/AlO$_{x}$(8nm) heterostructures. The initial and final experimentally measured Hall resistance (R$_{xy}$) before and after the application of a current pulse are shown in the middle with (a) a cyan and a green square (b) a yellow and a violet circle, respectively. The evolution of the Hall resistance with magnetic field is shown in red as a reference. The corresponding domain configurations before and after the application of electric pulse are shown schematically on the left and right, respectively. $``Up"$ and $``down"$ domains are shown schematically with triangles with a green and violet dots, respectively. Under a zero magnetic field, post demagnetization due to the current pulse, domains with both $``up"$ and $``down"$ magnetization nucleate in equal proportions, resulting in zero anomalous Hall resistance, shown in (a). In contrast, when a finite cooling field of $+0.2$T is applied during the switching process, all domains align in the $``up"$ direction, leading to a significant anomalous Hall resistance, shown in (b). Switching behavior as a function of (c) pulse amplitude with the pulse width fixed at 12ms and (d) pulse width with the pulse amplitude fixed at 100mA. The transient temperature during the application of pulse is estimated by measuring the longitudinal resistance (R$_{xx}$) \cite{suppl}. R$_{xx}$ is measured directly by an NI-DAQ device for low voltages, shown in green, and through a divider circuit for higher voltages, shown in blue. }
\label{fig:Switch_AHE}
\end{figure*}
%%%%%%%%%%%%%%%%%%%%%%%%%%%%%%%%%%%%%%%%%%%%%%%%%

%%%%%%%%%Fig.5%%%%%%%%%%%%%%%%%%%%%%%%%%%%%%%%%%%%%%
\begin{figure}
\includegraphics[width=1\columnwidth]{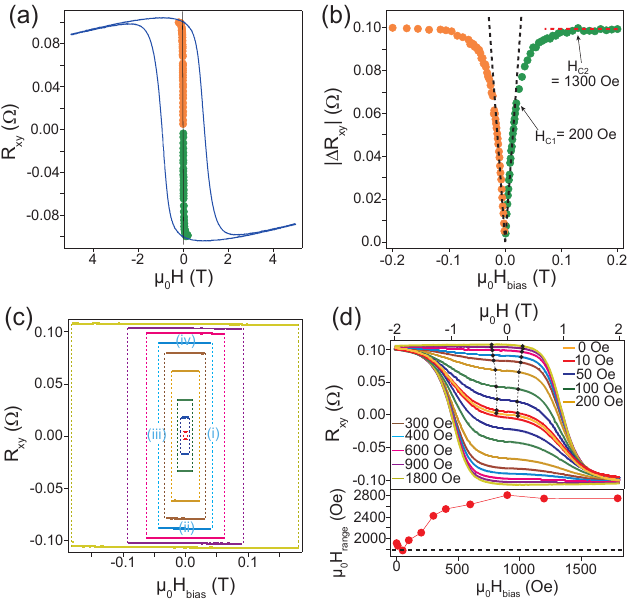}
\caption{ \textbf{Multiple-stable  anomalous Hall resistance memory states in \ms\/.} (a) Anomalous Hall resistance ($R_{xy}$) states after application of electrical pulse under different bias fields in Ta(11nm)/\ms\/(90nm)/Ta(11nm)/AlO$_{x}$(8nm) heterostructures. Results with positive and negative bias fields are shown in green and orange, respectively. The hysteresis curve of the AHE as a function of magnetic field is shown in blue for reference. (b) The change in $R_{xy}$ due to electrical switching under a bias field w.r.t the initial state. The initial state is obtained by applying an electric pulse with zero bias field that brings $R_{xy}$ to zero (see discussion in the text). The black dashed lines are linear fits. The red dashed line in (b) shows saturation behavior for fields above 1300 Oe. (c) Symmetrical switching behavior for positive and negative bias fields shown for different magnitudes of the bias field. The steps involved in obtaining a resistance loop are shown for a particular bias field, as described in the main text, in blue.(d) Evolution of the AHR memory states with magnetic field. The field range over which the AHR remains within $\pm 0.0021 \Omega$ of its value at zero field (see text) is shown with a black dot-dash line. The estimated stability field ranges as a function of bias fields is shown at the bottom. The dashed line in black corresponds to 1769 Oe, the minimum value obtained for the field range. In all cases, a single pulse of 100 mA and of duration 12 ms is applied for electrical switching.}
\label{fig:Switch_THE}
\end{figure}
%%%%%%%%%%%%%%%%%%%%%%%%%%%%%%%%%%%%%%%%%%%%%%%%%%

%%%%%%%%%%%%%%%%%%%%%%%%%%%%%%%%%%%%%%%%%%%%%%%%%%

%The nea in-plane lattice parameters $a$ $\&$ $b$ in our \ms\/ thin films were found to be slightly larger compared to the bulk single crystals with $a$ = 5.608 Å, $b$ = 5.555 Å $\&$ 

\end{document}